\documentclass[twocolumn,showpacs,preprintnumbers,amsmath,amssymb]{revtex4}

\usepackage{graphicx}
\usepackage{dcolumn}
\usepackage{bm}

\begin{document}



\title{
Microwave-induced suppression of dissipative conductivity
and  its\\ Shubnikov -- de Haas oscillations in two-dimensional electron systems:\\
Effect of dynamic electron localization
}

\author{Victor~Ryzhii* 
}
\address{Computer Solid State Physics Laboratory, University of Aizu,
Aizu-Wakamatsu 965-8580, Japan\\}

\date{\today}

\begin{abstract}
%

We present a model for microwave photoconductivity
in two-dimensional electron systems (2DESs) in a 
magnetic field
at the microwave frequencies
lower that the electron cyclotron frequency when
the intra-Landau level (LL) transitions dominate.
Using this model, we  explain the effect of decrease in 
the 2DES dissipative conductivity (and resistivity)
and smearing of its Shubnikov -- de Haas oscillations by microwave radiation
observed recently~\cite{1,2}. 
The model invokes the concept
of suppression of elastic impurity scattering of electrons
by the microwave electric field. We calculated the dependence
of the 2DES conductivity associated with  intra-LL
transitions as a function of the radiation and cyclotron frequencies
and  microwave power. The obtained dependences are consistent with
the results of recent experimental observations~\cite{1,2}.
\end{abstract}


\maketitle

\section{Introduction}

Recently, the experimental observations
of rather strong effect of relatively low frequency microwave
radiation (the radiation frequency $\Omega \lesssim \Omega_c$ or
$\Omega \ll \Omega_c$, where $\Omega_c$ is the electron cyclotron frequency)
on the dissipative conductivity and resistivity of two-dimensional
electron systems (2DESs) in GaAs/AlGaAs heterostructures were reported~\cite{1,2}. 
As shown, at sufficiently high microwave
powers, the dissipative conductivity (resistivity) of a 2DES
and its Shubnikov -- de Haas (SdH) oscillations are substantially suppressed.
The modulation of SdH oscillations
by microwaves was also reported by other authors (see, in particular,~\cite{3}).
The observed effect supplements the pattern of microwave-induced
transport phenomena in 2DESs related to the
microwave-induced zero-resistance and zero-conductance 
states revealed by Mani {\it et al.}~\cite{4} and Zudov {\it et al.}~\cite{5}
and extensively studied both experimentally and theoretically
last two years (for early predictions see, for example, Ref.~\cite{6}).

Leaving aside the hypothesis put forward in Ref.~\cite{1} to clarify
the effect, we provide an explanation associated with the suppression
of electron scattering on impurities with
intra-Landau level (LL) transitions by microwave radiation.
The main point is that the ac electric field 
of microwave (or optical) radiation
gives rise to spatio-temporal oscillations
of the electron Larmor orbit center that leads to an additional spreading
of the wave function. As  a result,  
the matrix elements of  electron-impurity interaction,
the probability
of  electron elastic scattering with a displacement
of the orbit center, and therefore, the dissipative 
conductivity are decreased by microwave radiation 
even without  absorption or emission of real photons. 

The effect of dissipative conductivity
suppression 
was predicted and assessed theoretically for  3DESs in magnetic field 
many years ago~\cite{7}. Similar effect of radiation-induced
suppression of 
electron transitions between two spatially separated states
(in neighboring quantum wells), called the dynamic electron
localization,
 was discussed and experimentally observed
by Keay {\it et al.}~\cite{8}.
The possible role of the effect of dynamic electron
localization in 2DESs in a magnetic field
was mentioned in Ref.~\cite{9}.
In this paper, we present a theoretical  model for this effect
and assess it.  It is demonstrated that invoking
the mechanism under consideration one can explain main features
of the experimental results.

\section{Electron transport due to intra-LL electron
transitions.}

The dissipative electron transport in the direction
parallel to  the electric field and perpendicular to the magnetic field
is due to hops of  electron  orbit centers caused by
scattering processes. These hops result in a change in the electron
potential energy $\delta \epsilon = - F\delta \rho$,
where  $F = - eE$ is the dc electric force,
$E$ is the dc electric field, and 
$\delta \rho = \rho_k - \rho_{k}^{\prime}$ is the displacement of 
the electron orbit center. 
When electron scattering on impurities 
dominates and  LL broadening is insignificant,
the displacement of  electron orbit center caused by
scattering with the electron  transition 
from the state  $(N, \rho_k)$ to
the state $(N^{\prime}, \rho_{k^{\prime}})$ 
one can  found
using the energy conservation law (disregarding spin-flip
 effects):
\begin{equation}\label{eq1}
F\delta \rho = \Lambda\hbar\Omega_c,
\end{equation}
where $\Lambda = N^{\prime} - N$ and $\hbar$
is the reduced Planck constant..
The matrix element, ${\cal M}_i$, of the  electron
transitions due to impurity scattering
and, consequently,
the probability, ${\cal W}$, of the scattering processes in question 
are determined by the overlap of  electron
wave functions (6) corresponding to two spatially separated states
(i.e., with different coordinates of the electron orbit center $\rho_k$)
and, therefore,  ${\cal W} = {\cal W}(\delta\rho)$ 
steeply decreases at sufficiently large $\delta\rho$.
Due to this,
 the pertinent contribution to the dissipative current is
\begin{equation}\label{eq2}
\delta J_D \propto - e\delta \rho {\cal W}(\delta\rho). 
\end{equation}

Considering Eqs.~(1) and (2) and taking into account
electron-impurity interactions in the Born approximation, 
one can arrive at the following
formula for the dissipative current 
which is equivalent to that
obtained by Tavger and Erukhimov~\cite{10}:
$$
J_D \propto 
 \nu_i\sum_{\Lambda}(f_N - f_{N+\Lambda})\biggl|A_{N,\Lambda}^{(i)}
\biggl(\frac{\hbar\Omega_c}{eEL}\biggr)\biggr|^2
$$
\begin{equation}\label{eq3}
\times
\biggl(\frac{\Lambda\hbar\Omega_c}{eEL}\biggr)
\exp \biggl[- \frac{1}{2}\biggl(\frac{\Lambda\hbar\Omega_c}{eEL}\biggr)^2\biggr].
\end{equation}
Here $\nu_i$ is the frequency characterizing  electron-impurity
collisions, $L$ is the quantum magnetic length,
$A_{N,\Lambda}^{(i)}$ is  determined
by the matrix elements of the impurity potential,
$f_N = [\exp(N\hbar\Omega_c/T - \zeta_F) + 1]^{-1}$
is the Fermi distribution function, $\zeta_F = \varepsilon_F/T$,
and $\varepsilon_F$
is the 2DES Fermi energy reckoned from
the lowest LL and normalized by temperature $T$.

In a nondegenerate 2DES in moderate electric fields
$E < E_c = \hbar\Omega_c/eL$ ( so that
$eEL < \hbar\Omega_c$),
 the transitions only 
between the neighboring
($\Lambda = 1$)  LLs with small indices are efficient. In this case,
Eq.~(3) yields for the dissipative conductivity the following expression:
\begin{equation}\label{eq4}
\sigma_D = \frac{J_D}{E} \propto 
 \nu_i
\exp \biggl[- \frac{1}{2}\biggl(\frac{E_c}{E}\biggr)^2\biggr].
\end{equation}
As pointed out, the exponential electric-field dependences 
given by Eqs.~(3) and (4)
are due to inter-LL transitions which become crucial under sufficiently
strong net dc electric  field. 
Such inter-LL transitions
can be called the Zener tunneling
transitions between LLs~\cite{11}. If the Zener tunneling
between LLs is due to  resonant transitions via impurity levels~\cite{12},
the calculated dissipative current-voltage characteristic
remains exponential.
In a degenerate 2DES with a large filling 
factor $N_F = \varepsilon_F/\hbar\Omega_c$,
the electron inter-LL transitions with $\delta\rho 
\sim L_F = L\sqrt{2N_F + 1}$ can provide the main contribution
to the dissipative current (see, for example,~\cite{11}). 
In this case, inter-LL electron transitions
become crucial
and 
a substantial increase in the dissipative conductivity
occurs when $E \gtrsim E_c^{(F)} = \hbar\Omega_c/L_F \simeq E_c/\sqrt{2N_F}$.

Equation~(4) yields a non-analytic electric-field
dependence which corresponds to $\sigma_D$ vanishing in the limit of weak
electric field~\cite{10}. 
However, nonanalytical dependences, like that given by
Eq.~(4), are  valid only if $|E| \gg E_b = \hbar\Gamma/eL,\, E_b^{(F)} = 
\hbar\Gamma/eL_F,$,
where $\Gamma$ is the LL broadening.
In relatively weak net dc electric fields,
the LL broadening becomes crucial~\cite{13}.
The dissipative conductivity of 2DESs in a magnetic field
associated with  impurity scattering 
under the assumption that  LL broadening
is due to electron-electron interactions 
($\Gamma \simeq \Gamma_{ee} > \Gamma_i \sim \nu_i$)
was considered in Ref.~\cite{13}.
The latter can be justified, in particular,
 in the 2DESs in which the electron sheet
concentration
$\Sigma$ is of the same order of magnitude as the sheet concentration
of remote donors (separated from the 2DES by sufficiently thick spacer). 
In strongly degenerate 2DESs, in which electron-electron scattering
is markedly weaken (see, for example,~\cite{14}), 
the contribution of impurity scattering to the LL net  broadening 
can be essential.
According to Eq.~(1),  
impurity scattering in weak electric field $|E| \ll E_c, E_c^{(F)}$
can
result only in the transitions with  rather large $|\delta \rho|$.
Due to very small overlap of the electron
wave functions in the initial and final states, the probability
of these transitions is exponentially small. This leads to
an exponentially small contributions of 
inter-LL 
transitions to
the dissipative current in small dc electric fields.
However, in the range of weak electric fields,
the dissipative conductivity can be associated with
electron  scattering on impurities within LLs provided
that LLs have a finite width. 
This implies that although electron-electron interaction does not
change the total momentum of the electron system and, hence, does not directly lead
to the dissipative conductivity, it may mediate the momentum transfer 
to the scatterers and, 
therefore strongly affect  electron transport phenomena
(see, for example, Ref.~\cite{15}). 
Invoking the mechanism of intra-LL impurity scattering
mediated by electron-electron collision, one can understand
the association of the dissipative current 
with the Joule heating of 2DES. 
Indeed, in the case of the mechanism in question,
the change in the electron potential energy associated with
the electron orbit centers displacements in the direction of the
electric force is compensated by an increase of the kinetic energy
(heating) of  all participated electrons.
The problem of Joule heating in 2DES in a magnetic field, 
to the best of our knowledge,
is not resolved in the framework of model considering
 solely impurity scattering (even beyond the Born approximation).
There is no such a problem in the case of 2DES in a magnetic field  because
the energy acquired by an electron 
from the electric field due to the displacement of its Larmor orbit
center associated with impurity scattering
goes to an increase of the energy of the electron motion along the
magnetic field.

When $\Omega_c \gg \Gamma$, the values of
characteristic fields $E_b$ and $E_b^{(F)}$, on the one hand,
 and 
$E_c$ and $E_c^{(F)}$, on the other, are strongly different.
For example, for a AlGaAs/GaAs 2DES,
assuming $H = 0.2$~T,  $\Gamma = 10^{10}$~s$^{-1}$, and $N_F = 50$
one can obtain $E_b \simeq 1$~V/cm, $E_b^{(F)} \simeq 0.1$~V/cm,
 $E_c \simeq 60$~V/cm, and $E_c^{(F)} \simeq 6$~V/cm.\\

As follows from Ref.~\cite{13},
in the case $\Gamma <  T/\hbar \ll \Omega_c$, 
the dissipative current (dissipative conductivity)
in low dc electric fields
is given by
\begin{equation}\label{eq5}
J_D \propto E \nu_i\sum_N b_N^{(i)}
\biggl(- \frac{\partial f_N}{\partial \zeta_F}\biggr).
\end{equation}
 Here 
\begin{equation}\label{eq6}
\frac{\partial f_N}{\partial \zeta_F} =
\frac{\exp(N\hbar\Omega_c/T - \zeta_F)}
{[1 + \exp(N\hbar\Omega_c/T - \zeta_F)]^2}
\end{equation}
and $b_N^{(i)}$
is a  coefficient determining by the matrix elements of the impurity potential
which, in turn, depend 
on the LL index. The most crucial feature of the formula for the dissipative
current  is that the latter is determined by the effective frequency
of electron collisions with impurities and by the derivative of the
electron distribution function~\cite{16}.

Using Eq.~(5), the low-field dissipative conductivity
can be presented as
\begin{equation}\label{eq7}
\sigma_D \propto  \nu_i\sum_N
\frac{ b_N\exp(N\hbar\Omega_c/T - \zeta_F)}
{[1 + \exp(N\hbar\Omega_c/T - \zeta_F)]^2}.
\end{equation}
As follows from Eq.~(6), at $T < \hbar\Omega_c$,
the dissipative conductivity is a strongly oscillating function of
$\Omega_c$ and $\zeta$. This is the well-known effect
of SdH
oscillations in 2DES 
(see, for example, Ref.~\cite{16})
One can see from Eq.~(6) that $\sigma_D$ reaches maxima when
$N\hbar\Omega_c = \varepsilon_F$, whereas the minima of $\sigma_D$
correspond to $(N + 1/2)\hbar\Omega_c = \varepsilon_F$.
At the minima, 
\begin{equation}\label{eq8}
\sigma_D \propto  \exp\biggl(-\frac{\hbar\Omega_c}{2T}\biggr).
\end{equation}
Thus, despite the dissipative conductivity in the case of low electric
field under consideration is determined by  electron transitions
within LLs, it exhibits an activation behavior with  activation energy
$\varepsilon_A = \hbar\Omega_c/2$ which is
determined by the separation between LLs.
At sufficiently low temperatures, the dissipative conductivity
exhibits  giant SdH  oscillations
with its exponentially small values under the condition
of quantum Hall effect  $N\hbar\Omega_c < \varepsilon_F < (N + 1)\hbar\Omega_c$.\\

\section{Effect of microwaves on intra-LL electron transitions}

To calculate the contribution of impurity scattering of electrons in
the dc electric and magnetic fields in the presence of microwave ac electric filed
to the dissipative conductivity, one can start from
the following Hamiltonian:
$$
{\cal H}_{{\cal E}} = - \frac{\hbar^2}{2m}\biggl[\frac{\partial^2}{\partial \rho^2} 
+ \biggl(\frac{\partial}{\partial \xi}  + i\frac{e}{c\hbar}H\rho\biggr)^2\biggr]
+ eE\rho 
$$
\begin{equation}\label{eq9}
+ e{\cal E}(t)(e_{\rho} \rho + e_{\xi}\xi).
\end{equation}
Here  ${\cal E}(t) = {\cal E}\cos\Omega t$
is the ac electric field of microwave radiation, which is taken
into account in the dipole approximation, and $e_{\rho}$ and $e_{\xi}$
are the components of the microwave field complex polarization
vector.
The Hamiltonian~(9) leads to  quasi-stationary states
which are characterized by quasi-energies.
The probability of an electron transition
from  state $(N, \rho_k)$ to  state $(N^{\prime}, \rho_{k}^{\prime})$
due to the scattering on impurities accompanied with
the absorption of $M$ real photons is proportional
to 
\begin{equation}\label{eq10}
|{\cal M}_{N,\rho_k, N^{\prime},\rho_{k^{\prime}}}^{(M)}|^2 
= |J_M(a_{\Omega}Q)|^2
|{\cal M}_{N,\rho_k, N^{\prime},\rho_{k^{\prime}}}|^2 .
\end{equation}
Here ${\cal M}_{N,\rho_k, N^{\prime},\rho_{k^{\prime}}}$ 
is the matrix element of electron-impurity
interaction without microwave radiation,
$J_M(x)$ is the Bessel function, $a_{\Omega} = L_{\cal E}/L$
is the relative amplitude of electron orbit center classical
oscillations in the crossed
dc electric and magnetic fields under the ac microwave electric field, 
and $Q = L(k^{\prime} - k)$.
Thus, the matrix element and, consequently, the
probability of an electron transition
due to the scattering on impurities accompanied with
the absorption or emission of $M$ real photons is proportional
to additional factor $|J_M(a_{\Omega}Q)|^2$.
In the case of circular polarization or 
when the microwave radiation is nonpolarized,
\begin{equation}\label{eq11}
L_{\cal E} = \frac {e{\cal E}}
{\sqrt{2}m\Omega^2}\frac {\Omega\sqrt{\Omega_c^2 + \Omega^2}}
{|\Omega_c^2 - \Omega^2|}. 
\end{equation}
The appearance of the
Bessel functions in  the scattering matrix
elements  is the result of calculation 
using the exact wave functions of electrons in both dc
and ac fields. This can be attributed to the processes
of absorption and emission of arbitrary number of virtual
photons in each process involving $M$ real photons.
Scattering processes in the presence of microwave ac electric field
are characterized by  the following two features
of  photon-assisted impurity scattering processes~\cite{7,9,17}.
First, $|{\cal M}_{N,\rho_k, N^{\prime},\rho_{k^{\prime}}}^{(M)}|^2$
  and, consequently the probability of
the processes involving $M$ real photons is 
not proportional to $|{\cal E}|^{2M}$ -
it is a more complex function of   ${\cal E}$ due to the Bessel
function dependence. 
Naturally, at low microwave powers (low ac electric fields),
$|J_1(a_{\Omega}Q)|^2 \propto a_{\Omega}^2Q^2$, and
the matrix element for single photon processes becomes
proportional to  $a_{\Omega}^2 \propto |{\cal E}|^2$, 
i.e., proportional to the microwave power $P_\Omega$.
Second, the probability of the impurity scattering
of electrons without the absorption of real photons ($M = 0$), i.e.,
elastic impurity scattering depends, nevertheless, on the microwave field.
This effect was theoretically studied in electron systems without
and with 
a magnetic field by different authors many years ago.
As pointed out above, microwave radiation can affect intra-LL elastic
impurity
scattering processes (involving no real photons) in 2DES changing
the effective frequency of electron collisions with impurities.
Such an effect can be explained by the following:
Increase in  the ac electric field  leads to an increase in
the amplitude of the electron Larmor orbit and, consequently, in a smearing
of the electron wave function. Due to this, the integral of the wave functions
with the same LL-index before and after the electron scattering and the
impurity potential  decreases
that leads to a decrease in the scattering probability. Indeed,
the matrix element
of impurity scattering without absorption or emission
of real photons ($M = 0$) is proportional to
$|J_0(a_{\Omega}Q)|^2$. 
As a result, generally $|{\cal M}_{N,\rho_k, N^{\prime},\rho_{k^{\prime}}}^{(0)}|^2
\neq |{\cal M}_{N,\rho_k, N^{\prime},\rho_{k^{\prime}}}|^2$
if $a_{\Omega} \neq 0$.
Consequently,
matrix element ${\cal M}_{N,\rho_k, N^{\prime},\rho_{k^{\prime}}}^{(0)}$
decreases (exhibiting
damping oscillations) with increasing $a_{\Omega}Q$
and even turns zero at  $a_{\Omega}Q \simeq 2.4$.
This implies that microwave radiation can effectively suppress
inter-LL impurity scattering and, consequently,
the hops of electron  orbit centers. The latter can be interpreted
as some kind of electron localization by
microwave radiation resulting in a decrease 
in the dissipative conductivity outside the cyclotron resonance and its harmonics
(compare with the effect of radiation-induced suppression of tunneling between quantum wells~\cite{8}).
The effect dynamic localization can be particularly important at relatively low
microwave frequencies $\Omega < \Omega_c$ when the dissipative
conductivity is mainly due to intra-LL impurity scattering
processes (the photon energy is insufficient for the inter-LL transitions).


Let the Fourier component of the impurity potential
to be  $V_q = V_i Q^s\exp (-d_iQ/L)$, where $Q = \sqrt{Q_x^2 + Q_y^2}$. Here $V_i$ is a constant (so that
$\nu_i \propto |V_i|^2$) and (a) 
for charged
remote impurities, $s = -1$ and
 $d_i$ is the spacing between the 2DES and the $\delta$-doped 
impurity layer, 
and (b) for short-range residual impurities $s = 0$ 
and $d_i = 0$.
After that, for coefficients $b_N$ in Eq.~(7) one can obtain~\cite{9}

$$
b_N = \frac{1}{\pi}\int dQ_x\,dQ_y Q_y^2Q^{2s}J_0^2(\xi_{\Omega}Q)
$$
$$
\times 
\exp\biggl(- \beta Q - \frac{Q^2}{2}\biggr) 
\biggl[L_N^0\biggl(\frac{Q^2}{2}\biggr)\biggr]^2 
$$
\begin{equation}\label{eq12}
= \int_0^{\infty}dQ  J_0^2(a_{\Omega}Q) Q^{3 + 2s}\exp\biggl(- \beta Q - \frac{Q^2}{2}\biggr) 
\biggl[L_N^0\biggl(\frac{Q^2}{2}\biggr)\biggr]^2,
\end{equation}
where $\beta = 2d_i/L$.  
At $s = -1$ and
$\beta \ll 1$ in the absence of microwave radiation ($a_{\Omega} = 0$),
from Eq.~(12) one obtains $b_N = b_N^0$ with 
\begin{equation}\label{eq13}
b_N^0 = \int_0^{\infty}dQQ \exp\biggl(- \frac{Q^2}{2}\biggr) 
\biggl[L_N^0\biggl(\frac{Q^2}{2}\biggr)\biggr]^2 = 1.
\end{equation}
For the case $s= 0$ and $\beta = 0$, instead of Eq.~(12), one obtains
$$
b_N^0 = \int_0^{\infty}dQQ^3\exp\biggl(- \frac{Q^2}{2}\biggr) 
\biggl[L_N^0\biggl(\frac{Q^2}{2}\biggr)\biggr]^2
$$
\begin{equation}\label{eq14}
= 2(2N + 1).
\end{equation}
The latter formula implies that in the case of
short-range impurity scattering, the contribution
of the transitions within the $N$th LL is proportional
to the square of the electron Larmor orbit radius $R_N = \sqrt{2N + 1}L$.
Comparing Eqs.~(13) and (14), one can see that the short-range impurity
scattering provides a larger contribution to the dissipative
conductivity by factor $2(2N + 1)$ (for the same impurity concentrations).
An increase in the spacer thickness $d_i$, i.e., an increase
of parameter $\beta$ lead for an additional
decrease in the contribution of scattering on charged impurities.
This is consistent with the previous conclusions~\cite{9,17}.

At not too high microwave powers and not too close to
the cyclotron resonance, one can expand the Bessel function
in the right-hand side of Eq.~(12). As a result, for the cases
of electron scattering on charged impurities ($s = -1$, $\beta \ll 1$)
and short-range scatterers ($s = 0$)
we arrive, respectively, at 
$$
b_N \simeq \int_0^{\infty}dQQ\biggl(1 - \frac{1}{2}\xi_{\Omega}^2 Q^2\biggr)
\exp\biggl(- \frac{Q^2}{2}\biggr) 
\biggl[L_N^0\biggl(\frac{Q^2}{2}\biggr)\biggr]^2
$$
$$
= 1 - 
\frac{1}{2}a_{\Omega}^2\int_0^{\infty}dQQ^3\exp\biggl(- \frac{Q^2}{2}\biggr) 
\biggl[L_N^0\biggl(\frac{Q^2}{2}\biggr)\biggr]^2
$$
\begin{equation}\label{eq15}
= 1 - (2N + 1)a_{\Omega}^2.
\end{equation}
$$
b_N \simeq \int_0^{\infty}dQQ^3
\biggl(1 - \frac{1}{2}a_{\Omega}^2 Q^2\biggr)\exp\biggl(- \frac{Q^2}{2}\biggr) 
\biggl[L_N^0\biggl(\frac{Q^2}{2}\biggr)\biggr]^2 
$$ 
$$
= 
2(2N + 1) - \frac{1}{2}\xi_{\Omega}^2 \int_0^{\infty}dQQ^5\exp\biggl(- \frac{Q^2}{2}\biggr) 
\biggl[L_N^0\biggl(\frac{Q^2}{2}\biggr)\biggr]^2 
$$
\begin{equation}\label{eq16}
= 2(2N + 1) - 4(3N^2 + 3N + 1)a_{\Omega}^2 
\simeq 4N(1 - 3Na_{\Omega}^2).
\end{equation}
 Assuming that the Fermi energy significantly exceeds the splitting
between LLs (large filling numbers), so that
the LL with large 
$N$ are most important, Eq.~(16) can be rewritten as
$$
b_N \simeq b_N^0 \biggl[1 - 2\biggl(\frac{3N^2 + 3N + 1}{2N + 1}\biggr)a_{\Omega}^2 \biggr] 
$$
\begin{equation}\label{eq17}
\simeq  b_N^0 
(1 - 3Na_{\Omega}^2).
\end{equation}

Considering Eqs.~(13) - (17), we obtain
\begin{equation}\label{eq18}
b_N \simeq b_N^0
\biggl[1 - k(s,\beta)N \biggl(\frac{2\pi\alpha P_{\Omega}}{m\Omega^3}\biggr)\,
{\cal F}(\Omega_c/\Omega)\biggr],
\end{equation}
where
$\alpha = e^2/\hbar c \simeq 1/137$,
 as follows from Eqs.~(16) and (17), 
$k(-1, 0) \simeq 2$ and  $k(0,0)  \simeq 3$, and 
${\cal F }(\omega_c) = \omega_c^2(\omega_c^2 + 1)/(\omega_c^2 - 1)^2$.
Using Eqs.~(7) and (18) and taking into account
that the main contributions to the dissipative conductivity
are due to intra-LL transitions within several LLs in the vicinity
of the Fermi energy (i.e., within the LLs 
with $N \simeq \varepsilon_F/\hbar\Omega_c$), we obtain
the following
formula for the dissipative conductivity associated with
intra-LL electron scattering on impurities
in the presence of microwave radiation:
\begin{equation}\label{eq19}
\sigma_D \simeq \sigma_D^0[ 1  - 
{\cal P}\,{\cal F}(\Omega_c/\Omega)],
\end{equation}
where 
$\sigma_D^0$ 
is the dark dissipative conductivity
exhibiting SdH oscillations
with varying $\Omega_c$. The introduced normalized
microwave power ${\cal P}$ is given by  
${\cal P} = P_{\Omega}/\overline{P_{\Omega}}$, where 
\begin{equation}\label{eq20}
\overline{P_{\Omega}} = \frac{m\Omega^4\hbar }{2\pi \alpha k(s, \beta)
\varepsilon_F}
\end{equation}
Setting the electron effective mass
$m = 6\times 10^{-29}$~g (GaAs), the electron sheet density
$\Sigma = 3\times 10^{11}$~cm$^2$ (as in Ref.~\cite{1}), and $s = 0$,  
at the microwave frequency
$f = \Omega/2\pi = 20$~GHz,
one can find $\overline{P_{\Omega}} \simeq 0.69$~mW/cm$^2$.
If $\Omega = const$, the normalized dissipative conductivity
(and, consequently, normalized resistivity) changes from  
$\sigma_D/\sigma_D^0 = 1 - {\cal P}/2\Delta$ at 
$\Omega_c/\Omega = 1 + \Delta \gtrsim 1$,
where $\Delta \ll 1$ (but $\Delta \gtrsim \Gamma/\Omega_c$), to   
$\sigma_D/\sigma_D^0 = 1 - {\cal P}$
at $\Omega_c \gg \Omega$, i.e., $\Delta \gg 1$. 
In particular, at $\Delta = 0.2$ and $\Delta = 2.0$,
Eq.~(19) yields $\sigma_D/\sigma_D^0 -  1 \simeq  - 18{\cal P}$ 
and  $\sigma_D/\sigma_D^0 - 1 \simeq  - 2.2{\cal P}$, respectively.
At elevated microwave powers, the dependence of  $\sigma_D$ on
${\cal P}$ becomes nonlinear.

\section{Results and discussion}

As follows from Eq.~(20), the microwave radiation
suppresses the dissipative conductivity. It decreases
both the averaged (smooth) value of the dissipative conductivity
and the amplitude of its oscillations.
Figures~1 - 3 demonstrate 
the dependences of the normalized dissipative conductivity 
of the 2DES dissipative conductivity 
associated with  the intra-LL 
electron transitions 
in dark conditions and under microwave radiation
calculated using the formulas obtained above on
ratio $\Omega_c/\Omega$. At fixed microwave frequency $\Omega$,
these dependences are  actually the
magnetic-field dependences.
In the vicinity of the cyclotron resonance, to take into account
the LL damping, function ${\cal F }(\omega_c)$ is replaced by
${\cal F }^{*}(\omega_c) = \omega_c^2(\omega_c^2 + 1)/
[(\omega_c^2 - 1)^2 + \gamma^2]$, where $\gamma = 2\Gamma/\Omega$.
The dissipative conductivity is normalized by its value
in the absence of microwave radiation 
at such a magnetic field that $\Omega_c = \Omega$.
It is assumed that the electron sheet concentration in the 2DES
is equal to $\Sigma = 3\times10^{11}$~cm$^{-2}$, so that
setting $\varepsilon_F = \pi\hbar^2\Sigma/m$, where $m$ 
is the electron effective mass,
at $T= 0.4$~K one obtains $\zeta_F \simeq 280$.
Apart from this, we set $\Gamma/\Omega_c = 0.02$,
 $s = 0$, and $d_i = 0$ (short-range impurity scattering)
in Figs.~1 and 2. In Fig.~3, the curves correspond to
the  cases of both short-range
and long-range impurity scattering. 
As seen from Fig.~1, at fixed microwave frequency $\Omega$, 
the degree of suppression  increases with increasing microwave power 
and decreases with increasing cyclotron
frequency, i.e., when the magnetic field becomes stronger. 
One needs to stress that in the vicinity of the cyclotron resonance
($\Omega_c/\Omega \gtrsim 1$), 
the contribution of the inter-LL photon assisted scattering
on impurities (which is not shown in Figs.~(1) - (3))
is substantial. However, this contribution rapidly decreases when
$\Omega_c/\Omega$ deviates from unity. 
\begin{figure}
\centerline{\includegraphics[width=75mm]{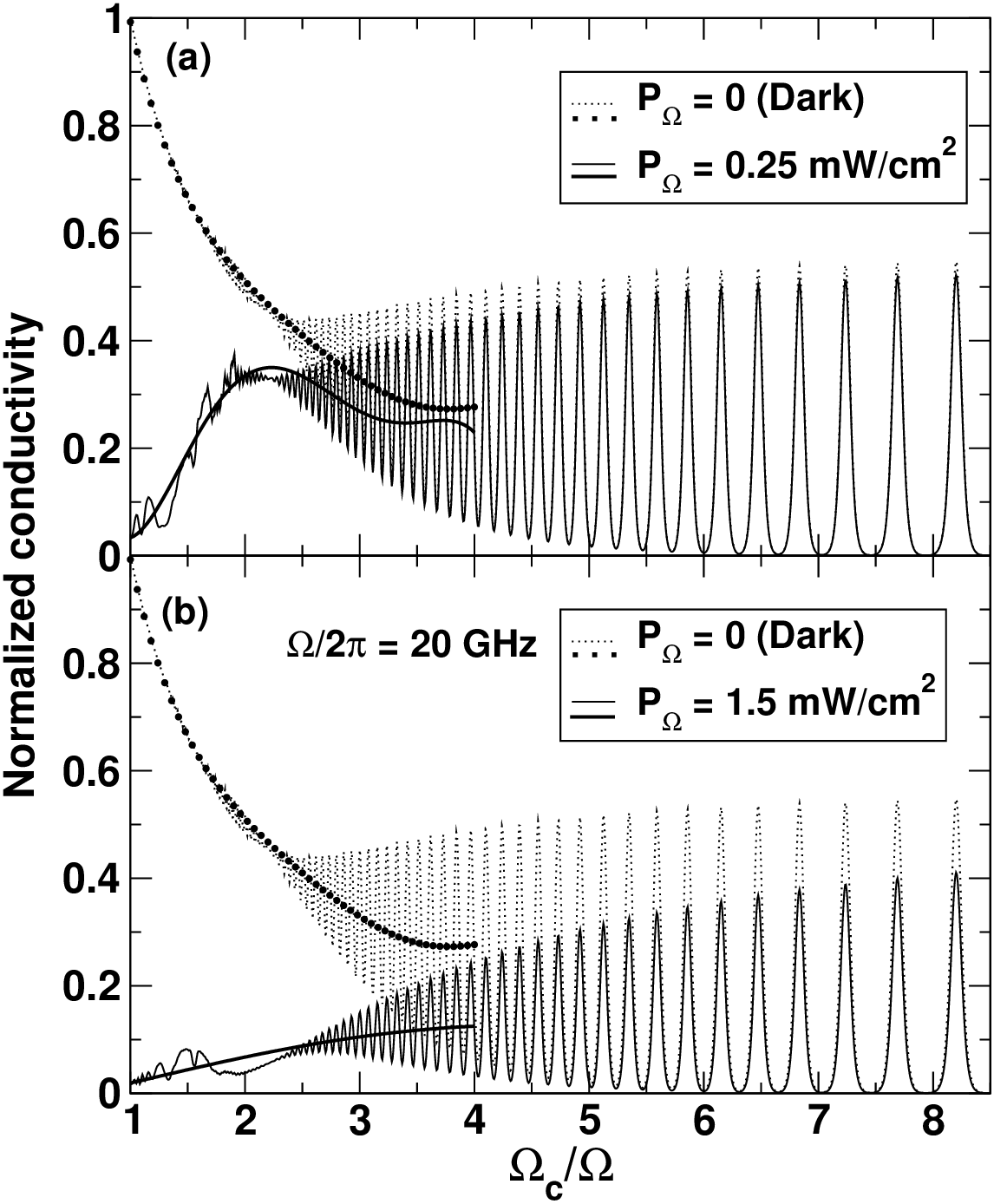}}
\caption{Magnetic-field dependences of  dissipative
conductivity
associated with intra-LL processes 
 normalized by dark dissipative conductivity
at $\Omega_c = \Omega$ at fixed microwave frequency 
and different microwave powers.
The conductivity averaged over   SdH oscillations
is show by bold dotted (dark conductivity) and solid (conductivity
in presence of microwave radiation) curves.
}
\end{figure}

\begin{figure}
\centerline{\includegraphics[width=75mm]{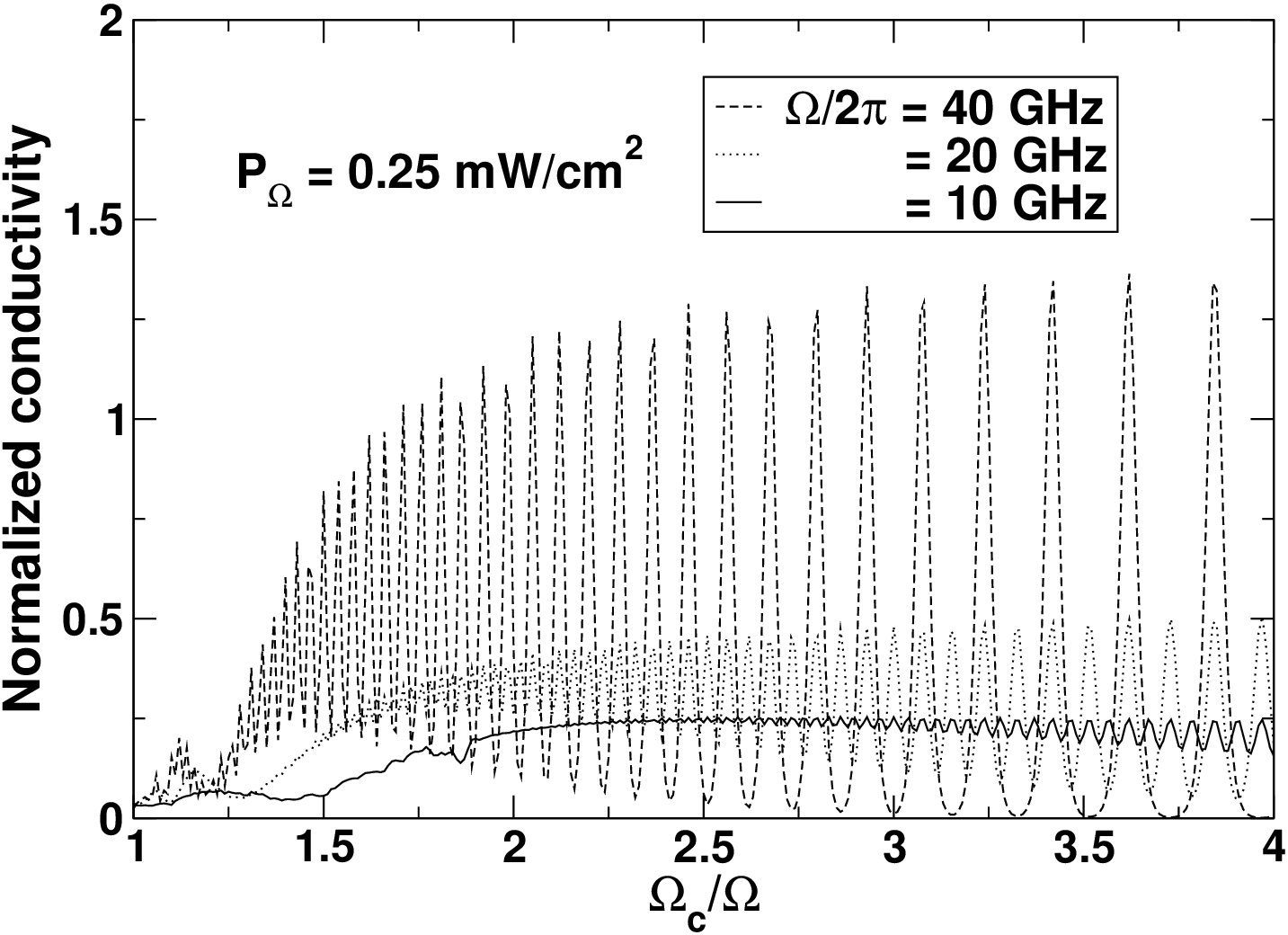}}
\caption{Normalized magnetic-field dependences of intra-LL conductivity
at different microwave frequencies.
}
\end{figure}

\begin{figure}
\centerline{\includegraphics[width=75mm]{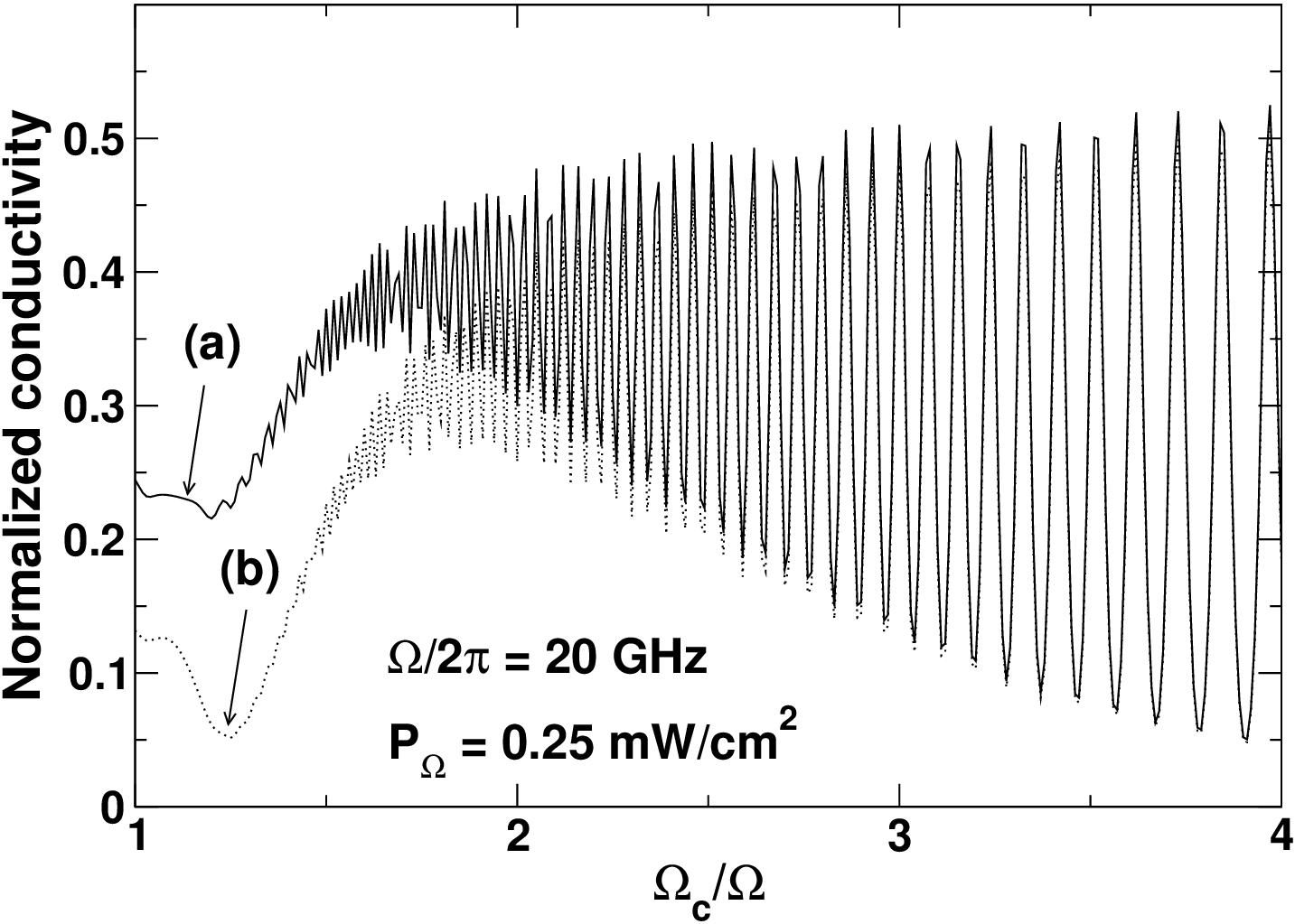}}
\caption{Normalized magnetic-field dependences of intra-LL  conductivity
calculated for
(a) long-range and (b) short-range impurity scattering mechanisms.
}
\end{figure}

Figure~2 shows that the effect of dissipative conductivity 
decrease and suppression of its
SdH oscillations  being rather strong
at relatively low microwave frequencies (for example, at $f = 10 - 20$~GHz) 
markedly
weakens with increasing microwave frequency. Indeed, as seen from Fig.~2,
at $f = 10 ~GHz$
the average conductivity is relatively small, 
and  its SdH oscillations are almost smeared.
However, an increase in the microwave frequency to $f = 40$~GHz
leads
to  marked increase of the average conductivity and recovery of
SDH oscillations. One can see that the amplitude of electron orbit center
oscillations in the limit $\Omega_c \gg \Omega$ is inversely proportional
to both $\Omega_c$ and $\Omega$. As follows from Eq.~(11), in this case,
$L_{\cal E} \propto {\cal E}/\Omega_c\Omega$~\cite{18}. Hence, an increase in
either $\Omega_c$ or $\Omega$ results in vanishing of the effect of microwave
radiation on intra-LL conductivity.
This is consistent with the experimental observations ~\cite{1,2}.

As found in the previous section,
the strength of the effect of microwave suppression under consideration
depends on the dominating impurity
scattering mechanism; the effect is more pronounced
when electrons are scattered primarily due a short-range interaction with
impurities. This is confirmed by the results of calculations
shown in Fig.~3. 

One needs to point out that Figs.~(1) - (3) show
the contribution to the 2DES dissipative conductivity
associated with  intra-LL electron transitions.
When $\Omega_c$ is moderately larger than
 $\Omega$,
photon-assisted inter-LL  scattering processes significantly
contribute to 
the conductivity. Since this contribution (at $\Omega_c \gtrsim \Omega$)
is positive and it is of the resonant nature, this resonant contribution
 surpasses the
contribution of intra-LL transitions in the immediate
vicinity of cyclotron resonance as  seen on the experimental plots~\cite{1}.
However, when $\Omega_c > \Omega$, the effect of dynamic suppression
of electron intra-LL scattering is dominant. At higher magnetic fields
when  $\Omega_c$ markedly exceeds  $\Omega$, the effect of microwaves on the
scattering being still dominant becomes weaker because of a 
decrease in the amplitude of
electron oscillations in the microwave ac electric field with increasing
magnetic field.

An increase in the electron temperature due to microwave heating
can affect the dissipative conductivity of 2DESs as well.
However, the electron heating leads to smearing
SdH
oscillations but it
does not change or slightly increases
(see, for example,~\cite{16} and references therein), 
the dissipative conductivity averaged over these
oscillations at least until the electron temperature remains small compared to
the Fermi energy.
However,  a marked decrease
in the dissipative conductivity averaged over SdH
oscillations was observed~\cite{1,2}. 
Due to this,  the heating mechanism appears to be irrelevant to
the effect observed experimentally in Refs.~\cite{1,2},
although this mechanism 
requires further theoretical analysis.

\section{Conclusions}

We proposed a model for low-frequency  microwave photoconductivity 
in  2DESs in a  magnetic field based on the concept
of dynamic localization of electrons by microwave ac electric field. 
We showed that
this model explains the following features of 
of microwave photoconductivity:\\
(1) Effective suppression of the dissipative conductivity 
and its SdH oscillations by microwave radiation
with  sufficiently high power in wide  magnetic field  range corresponding to
$\Omega_c/\Omega$ somewhat exceeding;\\
(2) Reincarnation of  SdH oscillations at sufficiently
large ratio $\Omega_c/\Omega$.

\section*{Acknowledgments}

The author is thankful to A.~Satou for assistance and
K.~von Klitzing and S.~Dorozhkin for stimulating comments.


\begin{references}
\bibitem{18}
At microwave frequencies much smaller than the cyclotron frequency,
the oscillations of electron orbit center are mainly due to its  Hall
drift in the ac electric field. In this case, the amplitude of such oscillations
$L_{\cal E} \sim  v_{\cal E}/\Omega$, where $v_{\cal E} = c{\cal E}/H \propto 
{\cal E}/\Omega_c$
is the Hall velocity in the crossed ac electric and dc magnetic fields and
$c$ is the speed of light. As a result, $L_{\cal E} \propto  
{\cal E}/\Omega_c\Omega$.
\end{references}
\end{document}